\documentclass[submission,copyright,creativecommons]{eptcs}
\providecommand{\event}{WIVACE 2013} 
\usepackage{breakurl}             
\usepackage{graphicx}

\title{A model of protocell based on the introduction of a semi-permeable membrane in a stochastic model of catalytic reaction networks}
\author{Roberto Serra \qquad\qquad Marco Villani
\institute{Department of Physics, Informatics and Mathematics\\
Modena and Reggio Emilia University}
\institute{ECLT\\
European Centre for Living Technology\\
S. Marco 2940, University of Ca' Foscari, Venice (IT)}
\email{rserra@unimore.it \qquad\qquad marco.villani@unimore.it}
\and
Alessandro Filisetti 
\institute{ECLT\\
European Centre for Living Technology\\
S. Marco 2940, University of Ca' Foscari, Venice (IT)}
\email{\quad alessandro.filisetti@unive.it}
\and
Alex Graudenzi \qquad\qquad Chiara Damiani 
\institute{Department of Informatics\\
Systems and Communication, University of Milan Bicocca}
\email{\quad alex.graudenzi@unimob.it \qquad\qquad chiara.damiani@unimib.it}
}

\def\titlerunning{A model of protocell}
\def\authorrunning{R. Serra et al.}
\begin{document}
\maketitle

\section*{Abstract}

The theoretical characterization of the self-organizing molecular structures emerging from ensembles of distinct interacting chemicals turns to be very important in revealing those dynamics that led to the transition from the non-living to the living matter as well as in the design of artificial protocells~\cite{Rasmussen2003,Serra:2006aa,Szostak:2001xw}. \\ 
In this work we aim at studying the role of a semi-permeable membrane, i.e. a very simple protocell description, in the dynamics of a stochastic model describing randomly generated catalytic reaction sets (CRSs) of molecules. \\
Catalytic reaction sets are networks composed of different molecules interacting together leading to the formation of several reaction pathways. In particular, molecules are representation of molecular species formed by a concatenation of letters (bricks) from a finite alphabet. \\
Reaction pathways are formed by two kinds of reaction only, namely condensation, where two molecular species are glued together in order to create a longer species and cleavage where a molecular species is divided in two shorter species, Figure 1, Filisetti et al.~\cite{Filisetti2011a,Filisetti:2010fk}. We present here only a brief summary of the principal characteristics and a description of the new features. \\
An assumption of the model is that each reaction needs a catalyst to occur; hence spontaneous reactions are neglected at this stage of the model. 

\begin{figure}[ht]
\begin{center}
\includegraphics[width=8cm]{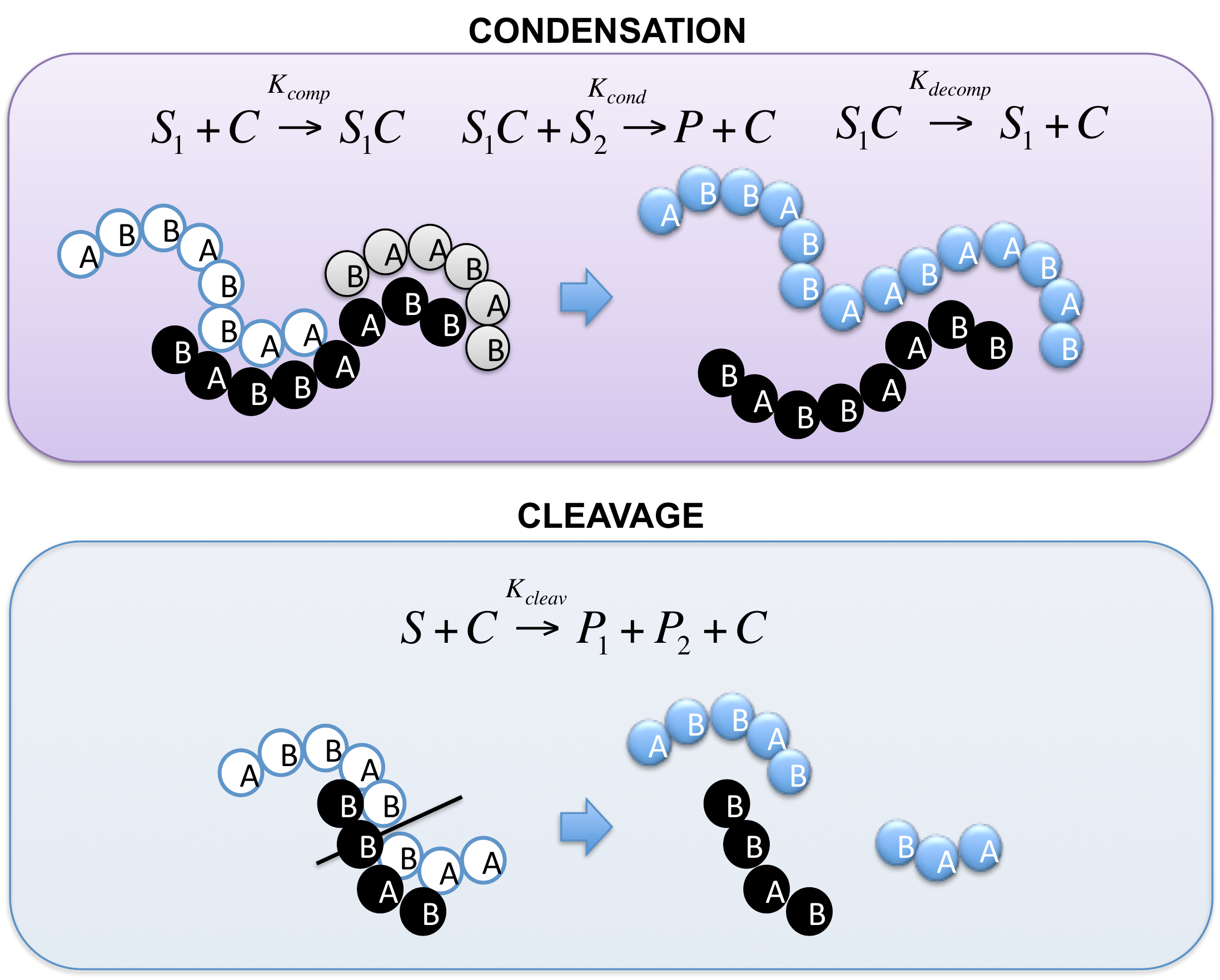}
\caption{A schematic representation of the two allowed reactions. TOP: Condensation reaction. The white and the grey chains stand for the substrates, the black chain represents the catalyst and the blue chain stands for the product. The reaction occurs in two steps, the formation of a temporary molecular complex and the release of the product. BOTTOM: Cleavage reaction. The white chain represents the substrate, the black chain stands for the catalyst and the blue chains are the two products of the reaction. }
\end{center}
\label{fig:schemino}
\end{figure} 

To create the reaction scheme, the molecular species involved in the reactions are randomly chosen at the beginning of the simulation according to a fixed probability \emph{p} that a random species catalyzes a random reaction. 
In accordance with both the species and the reactions created, in different realizations the same species can catalyze different reactions leading to the formation of different ÒchemistriesÓ, i.e. different possible artificial worlds. \\
Several models of catalytic reaction networks have been proposed. Some of them focus on the structural proprieties of the reaction graph~\cite{Farmer1986,Hordijk:2004vl,Kauffman:1986mi} while dynamical investigations in a continuous stirred-tank reactor (CSTR) have been introduced by Bagley et al.~\cite{Bagley1989}, Farmer et al.~\cite{Farmer1986,J.D.Farmer:1986ph}, Hordijk et al.~\cite{Hordijk2010,Hordijk2012a,Hordijk2012} and Vasas et al.~\cite{Vasas2012}. \\
We introduce in this model a semi-permeable membrane able to entrap all those molecular species longer than a certain threshold. \\
In order to reveal the role of the semi-permeable membrane we compare the dynamics of the same chemistry, i.e. the same reaction scheme, within a CSTR and a protocell (kinetic parameters are the same as well). \\
An important feature of the model proposed here is that in accordance with the reaction scheme new molecular species can be created by the dynamics. Each time that a new species is created, according to \emph{p} new reactions involving the possible interactions between the new species and the already present species are created. Thus, the system can grow in time increasing the number of molecular species and the number of reactions, Filisetti et al.~\cite{Filisetti2011a,Filisetti:2010fk}. \\
In order to consider the role of stochastic fluctuations due to the low-number-effects on the overall dynamics, we adopt a stochastic description of the system, thus dynamics are simulated using a hybrid version the well-known Gillespie algorithm~\cite{Gillespie:1977fv}.\\ 
The first remarkable difference between the dynamics within a CSTR and within a protocell concerns the overall mass of the system here represented as the total number of bricks present in the system. \\
In accordance with the nature of the system, a continuous incoming flux of nutrients and a continuous outgoing flux of all the molecular species, within a CSTR the overall mass reaches always a dynamical equilibrium. On the opposite, a protocell may never reach a dynamical equilibrium. In particular, while within a CSTR distinct simulations of the same chemistry show the same dynamical equilibrium, within a protocell the same chemistry reaches different dynamical states starting from the same initial conditions. 
A further important outcome is directly connected with the presence of the semi-permeable membrane. Some molecular species, that are expelled adopting a CSTR framework, remain entrapped within a protocell if longer than a certain threshold. This process leads to the formation of a pool of catalysts, no longer created by the system dynamics, responsible of the conservation, or of the dismantling, of certain reaction pathways. It is worthwhile to remark that the direct consequence of this phenomenon is the formation of different protocell compositions, a necessary requirement for the establishment of a population of different protocells competing for the same limited resources. \\
Finally, an important compositional remark concerns the maximum length of the species created by the dynamics. While within a CSTR, always observing the dynamics of the same chemistry, only species with a maximum length of 8 are observed, protocells are able to produce species up to length 11 indicating a greater capability to produce more complex compositions. \\
Further analysis are necessary for a better comprehension of the role played by a semi-permeable membrane on the dynamics of catalytic reaction networks.  

\nocite{*}
\bibliographystyle{eptcs}
\bibliography{biblio}
\end{document}